\documentclass[a4paper,10pt]{article}

\pdfoutput=1

\usepackage{jheppub}

\usepackage{graphicx}    
\usepackage{dcolumn}    
\usepackage{bm}             
\usepackage{amssymb}   
\usepackage{subfigure}
\usepackage{verbatim}
\usepackage{environ}

\usepackage[english]{babel}
\usepackage{xcolor}
\usepackage{graphicx}
\usepackage{amsmath}
\usepackage{dsfont}
\usepackage[makeroom]{cancel}
\usepackage{url}
\usepackage{hyperref}
\usepackage{slashed}

\newcommand{\secn}[1]{Section~\ref{#1}}

\newcommand{\beq}{\begin{eqnarray}}
\newcommand{\eeq}{\end{eqnarray}}

\newcommand{\eps}{\epsilon}

\newcommand{\zbar}{\bar{z}}
\newcommand{\barpartial}{\partial_{\zbar}}
\newcommand{\e}{\epsilon}
\newcommand{\bra}{\langle}
\newcommand{\ket}{\rangle}

\def\eq#1{Eq.~(\ref{#1})}

\preprint{\\ \rightline{CERN-TH-2021-063}}

\title{
Non-abelian infrared divergences on the celestial sphere
}

\author[a,b]{Lorenzo Magnea}

\affiliation[a]{Theoretical Physics Department, CERN, CH-1211 Geneva 23, Switzerland}
\affiliation[b]{Dipartimento di Fisica and Arnold-Regge Center, Universit\`a di Torino,\\
                 and INFN, Sezione di Torino, Via P. Giuria 1, I-10125 Torino, Italy}

\emailAdd{lorenzo.magnea@unito.it}

\abstract{
We consider the infrared factorisation of non-abelian multi-particle scattering amplitudes, 
and we study the form of the universal colour operator responsible for infrared divergences, 
when expressed in terms of coordinates on the `celestial sphere' intersecting the future 
light-cone at asymptotic distances. We find that colour-dipole contributions to the infrared 
operator, to all orders in perturbation theory, have a remarkably simple expression in these 
coordinates, with scale and coupling dependence factorised from kinematics and colour. 
Generalising earlier suggestions in the abelian theory, we then show that the infrared
operator can be computed as a correlator of vertex operators in a conformal field 
theory of Lie-algebra-valued free bosons on the celestial sphere. We verify by means
of the OPE that the theory correctly predicts the all-order structure of collinear limits, 
and the tree-level factorisation of soft real radiation.}

\begin{document}
\maketitle


\section{Introduction}
\label{Intro}

The problem of infrared divergences in gauge theory amplitudes and cross sections
has been studied for close to a century, starting with the seminal work of Bloch and
Nordsieck in QED~\cite{Bloch:1937pw}, and solutions involving an increasing degree
of generality and depth have been put forward across the decades. One class of 
proposed solutions is based upon the idea of carefully constructing well-defined
(and thus finite) observable quantities, in the spirit of the KLN theorem~\cite{Kinoshita:1962ur,
Lee:1964is}. Within this framework, one accepts that $S$-matrix elements are
ill-defined in massless theories, introduces an infrared regulator, and verifies
the cancellation of singularities for infrared-safe observables~\cite{Sterman:1978bi,
Sterman:1978bj}. Alternatively, one can try to tackle the infrared problem at the root, 
recognising that it originates from an inadequate definition of asymptotic states. 
This viewpoint led to the development of the method of coherent states, by 
means of which an infrared-finite perturbative $S$-matrix can be defined, both 
in QED~\cite{Kulish:1970ut} and for non-abelian theories~\cite{Giavarini:1987ts}. 
Remarkably, the study of infrared singularities for non-abelian gauge theories 
continues to this day to be a very active field of research, both for theoretical 
reasons, and in view of phenomenological applications. On the theoretical side, 
a general understanding of the all-order infrared factorisation properties of 
non-abelian multi-particle scattering amplitudes has been developed over 
the years~\cite{Mueller:1979ih,Collins:1980ih,Sen:1981sd,Sen:1982bt,
Collins:1989bt,Magnea:1990zb,Catani:1998bh,Sterman:2002qn,Dixon:2008gr,
Becher:2009cu,Gardi:2009qi,Becher:2009qa,Gardi:2009zv,Gardi:2013ita,
Feige:2014wja,Erdogan:2014gha,Ma:2019hjq}, and the relevant soft and 
collinear anomalous dimensions have been computed to very high orders, 
uncovering interesting new mathematical structures~\cite{Gardi:2010rn,
Mitov:2010rp,Gardi:2011wa,Gardi:2011yz,Vladimirov:2014wga,
Vladimirov:2015fea,Almelid:2015jia,Henn:2016jdu,Almelid:2017qju,Henn:2019swt,
vonManteuffel:2020vjv,Agarwal:2020nyc,Agarwal:2021zft,Agarwal:2021him}. 
On the phenomenological side, a vast effort has been devoted to the construction 
of efficient algorithms for the subtraction of infrared singularities at high 
perturbative orders, in both  virtual corrections and real radiation contributions 
to cross sections of experimental interest at colliders. In this case, the most 
significant problem is the treatment of unresolved real radiation for complex 
observables, in the presence of intricate experimental cuts. Recent advances 
in this direction have been reviewed in Ref.~\cite{TorresBobadilla:2020ekr}.

All the developments briefly summarised above are based on standard, if advanced,
techniques of quantum field theory in four dimensions, including diagrammatic tools,
renormalisation group equations, Ward identities and effective field theories. In recent 
years, starting with Refs.~\cite{Strominger:2013lka,Strominger:2013jfa}, a remarkable 
novel formulation of the infrared problem has been introduced, based on the discovery 
of enhanced long-distance symmetries of theories involving massless particles. 
Within this framework, the infrared behaviour of scattering amplitudes is
determined by infinite-dimensional asymptotic symmetries of the massless 
theory under consideration, acting on the `celestial' sphere intersecting the future 
(or past) light cone at asymptotic distances. For gravity, these symmetries have
been known for a long time~\cite{Bondi:1962px,Sachs:1962wk}, while in the case 
of gauge theories they take the form of `large' gauge transformations with non-trivial 
action at null infinity~\cite{Lysov:2014csa,He:2014cra,He:2015zea,Adamo:2015fwa,
Strominger:2015bla,Gabai:2016kuf,Cheung:2016iub,Nande:2017dba}. In this 
context, soft factorisation theorems for tree-level radiative amplitudes, and their 
next-to-leading-power extensions, emerge as Ward identities of these asymptotic 
symmetries. The first few years of work in this fast-developing field have been 
reviewed in~\cite{Strominger:2017zoo}, where one can find ample references 
to the literature.  At the moment, much of the work done within this framework is 
focused on tree-level amplitudes for gauge theories and gravity (see, however, 
Refs.~\cite{Albayrak:2020saa, Gonzalez:2020tpi}, where one-loop effects are 
analysed), but the connection that has been uncovered between infrared properties 
of $d=4$ massless field theories and $d = 2$ conformal invariance on the celestial 
sphere is intriguing, and represents a radically innovative point of view on an old 
problem. An important new tool introduced in this context has been the use of 
Mellin transforms of scattering amplitudes~\cite{Pasterski:2016qvg,Pasterski:2017kqt}, 
which are constructed to have simple transformation properties under the conformal
group acting on the celestial sphere, while from a four-dimensional viewpoint
they are boost eigenstates. A rapidly growing body of literature (see, for 
example, Refs.~\cite{Pasterski:2017ylz,Schreiber:2017jsr,Stieberger:2018onx,
Himwich:2019dug,Fan:2019emx,Pate:2019mfs,Nandan:2019jas,
Fotopoulos:2019tpe,Ebert:2020nqf,Arkani-Hamed:2020gyp,Fan:2021isc}) 
is exploring the properties of these {\it celestial amplitudes}, including 
detailed comparisons with known soft and collinear limits. Interestingly,
related ideas have also started to be applied in a more phenomenological 
context~\cite{Neill:2020bwv}.

In this paper, we will take a different viewpoint: instead of considering the full
multi-particle scattering amplitude, we will take advantage of its factorisation
properties in the infrared, and focus on the universal infrared colour operator
responsible for generating all soft and collinear singularities. Colour correlations
for this operator are well understood to all orders in perturbation theory, and
they originate from correlators of straight, semi-infinite Wilson lines, emerging 
from the hard scattering and pointing in the light-like directions defined by the 
momenta of the hard particles participating in the scattering. The role of scale
and conformal invariance in establishing the form of this infrared operator was
already pointed out in Refs.~\cite{Gardi:2009qi,Becher:2009qa}, and the 
conformal properties of the infrared factor were instrumental in the determination
of the three-loop infrared anomalous dimension by a bootstrap approach in 
Ref.~\cite{Almelid:2017qju}. More recently, a very important observation
in this direction was made in Ref.~\cite{Kalyanapuram:2020epb}: examining 
the abelian case, Kalyanapuram noticed that the one-loop soft factor 
for QED-like theories is very closely connected to a correlator of vertex 
operators for free bosons on the celestial sphere. Here, we will pursue 
this remarkable connection, and propose an extension to the non-abelian case.

At the outset, such a generalisation appears to be difficult to achieve, for 
several reasons: first of all, in QED the exponentiation of the infrared factor is 
essentially one-loop exact (as it is in gravity~\cite{Weinberg:1965nx}), while, in 
a non-abelian theory, intricate corrections emerge at all perturbative orders; 
next, as a consequence, the scale dependence of the non-abelian soft operator 
is non-trivial, and possibly entwined with colour correlations; finally, and more 
importantly, the non-abelian infrared factor is a colour operator and not just a 
number. These obstacles not withstanding, we will be able to propose a 
generalisation of the results of Refs.~\cite{Kalyanapuram:2020epb} to  
non-abelian massless gauge theories, in terms of a conformal theory 
of {\it Lie-algebra-valued} free bosons on the celestial sphere. 

We will begin in \secn{nonabsad} by providing a bird's-eye view of our current 
understanding of infrared divergences for massless non-abelian scattering amplitudes, 
focusing our attention on colour correlations, which arise exclusively from 
soft gauge boson exchanges. In \secn{sphecol}, concentrating on colour
correlations with a dipole structure, which are understood to all orders, 
we will translate known results in terms of variables defined on the
celestial sphere. Remarkably, we will find that all dependence on the
gauge coupling, on the scale, and on the dimensional regulator factorises
from kinematic and colour correlations, and can be collected in a single 
universal function which is well-known in perturbative QCD, giving, among 
other things, the all-order perturbative expression for the gluon Regge 
trajectory in high-energy scattering~\cite{Korchemsky:1993hr,Korchemskaya:1994qp,
Korchemskaya:1996je,DelDuca:2011xm,DelDuca:2011ae}; dependence
on colour and kinematics is then strikingly simple, and naturally expressed
in terms of distances on the celestial sphere. In \secn{matfrebo}, we introduce
a Lie-algebra-valued free-boson conformal theory on the celestial sphere.
Vertex operators in this theory are colour matrices, in representations 
corresponding to the hard particles participating in the scattering. That
notwithstanding, they have a well-defined conformal weight, and they
form well-behaved correlation functions, which could be seen as
colour-kinematic duals~\cite{Bern:2008qj,Bern:2019prr} of the integrands 
of bosonic string amplitudes. These correlators exactly reproduce the 
all-order expression for infrared divergences of gauge amplitudes with 
dipole colour correlations.

In order to further test the correspondence between the conformal theory and 
the gauge theory, we note the presence in the celestial theory of a Lie-algebra-valued 
conserved Noether current, whose OPE with vertex operators reproduces the 
tree-level soft-gluon current responsible for soft real radiation in the gauge theory.
This provides strong evidence for identifying this Noether current with the
Kac-Moody current constructed in Ref.~\cite{Strominger:2013lka,He:2015zea}
by considering asymptotic limits of four-dimensional gauge fields. Further, we 
use the conformal OPE of vertex operators to study collinear limits of the infrared 
factor, and we recover the expected expression for the splitting anomalous 
dimension for dipole correlations~\cite{Catani:2003vu,Becher:2009qa}, once 
again an all-order result. In \secn{Open}, we discuss some of the many open 
questions that our proposal leaves open at this stage. In particular,
we discuss the matching between the celestial coupling and the gauge-theory
coupling factor, noting that it would be of great interest to develop an
understanding of the role played by scale dependence in the correspondence
between the celestial and the four-dimensional theory. We emphasise 
that, in our view, the theory as presented here is incomplete, and we 
expect that a generalisation, or possibly a deformation, of the conformal
theory we propose, should be able to predict, and indeed compute,
the quadrupole correlations which arise in the gauge theory starting 
at three loops, and are explicitly known at that order~\cite{Almelid:2015jia,
Henn:2016jdu,Almelid:2017qju}. We conclude in \secn{Perspe} summarising 
our results and suggesting several open lines for future research.


\section{Non-abelian soft anomalous dimensions: a primer}
\label{nonabsad}

The structure of infrared divergences for on-shell scattering amplitudes in massless 
non-abelian gauge theories is understood in remarkable generality. To summarise
what is known, consider an $n$-point amplitude in dimensional regularisation, which 
we write as
\beq
  {\cal A}^{\, a_1 \ldots a_n}_n \bigg( \frac{p_i}{\mu}, \alpha_s (\mu), \e \bigg) \, ,
\label{amp}
\eeq
displaying the colour indices $a_i$ for each external particle. Note that we are allowing
for particles in different representations of the gauge group, so $a_i$ is not necessarily 
an adjoint index; furthermore, renormalisation has already been performed, and we
are working in $d = 4 - 2 \e > 4$, so that $\e < 0$, to regularise infrared singularities;
finally, we are not displaying polarisation indices, which will be irrelevant in what follows.
If desired, one can select a basis of colour tensors $c_M^{\, a_1 \ldots a_n}$ spanning
the space of allowed colour configurations for the process at hand: the amplitude in
\eq{amp} is a vector in that space, whose components in the chosen basis can be 
computed by suitable projections.

Infrared divergences arise from long-distance exchanges of virtual particles, and
they can be shown to factorise in the form~\cite{Sen:1982bt,Dixon:2008gr,Gardi:2009qi,
Becher:2009qa,Feige:2014wja,Ma:2019hjq}
\beq
  {\cal A}_n \bigg( \frac{p_i}{\mu}, \alpha_s (\mu), \e \bigg) \, = \, 
  {\cal Z}_n \bigg( \frac{p_i}{\mu}, \alpha_s (\mu), \e \bigg) 
  {\cal H}_n \bigg( \frac{p_i}{\mu}, \alpha_s (\mu), \e \bigg) \, ,
\label{factampl}
\eeq
where colour indices are understood, ${\cal  Z}_n$ is a universal colour operator 
generating all infrared poles in $\e$, and acting on a colour vector ${\cal H}_n$, 
which is process-dependent and finite as $\e \to 0$. The divergent factor ${\cal Z}_n$
satisfies a renormalisation group equation, which can be solved in the form
\beq
  {\cal Z}_n \bigg( \frac{p_i}{\mu}, \alpha_s (\mu), \e \bigg) \, = \, 
  P \exp \left[ \frac{1}{2} \int_0^{\mu^2} \frac{d \lambda^2}{\lambda^2} \,
  \Gamma_n \bigg( \frac{p_i}{\lambda}, \alpha_s (\lambda, \e) \bigg) \right] \, ,
\label{solZ}
\eeq
where $\alpha_s(\lambda, \e)$ is the $d$-dimensional running coupling, satisfying
\beq
  \lambda \frac{\partial \alpha_s}{\partial \lambda} \, \equiv \, \beta(\alpha_s, \e) \, = \, 
  - 2 \e \alpha_s - \frac{\alpha_s^2}{2 \pi} \, \sum_{k = 0}^\infty \bigg( \frac{\alpha_s}{\pi}
  \bigg)^{\! k} b_k \, .
\label{beta}
\eeq
\eq{solZ} features the central object of our discussion, the {\it infrared anomalous dimension
matrix}, which we denote by $\Gamma_n$. Note that the initial condition in \eq{solZ}
has been fixed in the infrared, by making use of the fact that $\alpha_s (\lambda = 0,
\e < 0) = 0$; note also that the matrix $\Gamma_n$ is finite as $\e \to 0$, and all
infrared singularities arise from the scale integration.

The matrix $\Gamma_n$ is clearly a fundamental object for perturbative gauge 
theories: for massless theories, it is fully known up to three loops for any $n$ and
for any gauge group and representation content~\cite{Almelid:2015jia}. What is most 
remarkable, however, is the fact that its all-order structure is strongly constrained by 
an underlying scale invariance of the infrared sector. In order to describe this structure, 
it is best not to write $\Gamma_n$ is a specific basis, but rather to adopt the 
basis-independent colour-operator notation of Refs.~\cite{Bassetto:1984ik,Catani:1996vz}. 
In this language, one introduces, for each hard particle $i$, $i = 1, \ldots, n$, a colour 
operator ${\bf T}_i$, which acts on the finite factor ${\cal H}_n$ as a generator 
of the gauge algebra in the appropriate representation\footnote{Thus, for 
example, for an outgoing quark one has ${\bf T}_i \to T^a_{\,\,ij}$, where $T^a$ 
is a generator in the fundamental representation of $SU(N)$, while for a gluon 
one has ${\bf T}_i \to - {\rm i} f^a_{\,\,c d}$, where $f^a_{\,\,c d}$ are the structure 
constants.}. The colour operators ${\bf T}_i$ carry an adjoint index, and satisfy
\beq
  \Big[ {\bf T}_i^a, {\bf T}_i^b \Big] \, = \, {\rm i} f^{ab}_{\,\,\,\,\, c} \, {\bf T}_i^c \, ,
  \qquad 
  {\bf T}_i \cdot {\bf T}_i \, \equiv \, {\bf T}_i^a {\bf T}_i^b \, \delta_{ab} \, = \, C_i^{(2)} \, ,
  \qquad
  \sum_{i = 1}^n {\bf T}_i \, = \, 0 \, , 
\label{propT}
\eeq 
where $C_i^{(2)}$ is the quadratic Casimir eigenvalue of the gauge algebra in the
representation of particle $i$, while the last equation enforces colour conservation, 
and must be interpreted as constraint on the action of the colour operators ${\bf T}_i$
on the finite factor ${\cal H}_n$, dictated by gauge invariance. Note that colour 
operators associated with different hard particles commute, since ${\cal Z}_n$ 
is an operator in the tensor product of the colour representations of the $n$
particles, and each ${\bf T}_i$ acts on a single factor in that product.

Let us now discuss the all-order structure of the infrared anomalous dimension matrix.
As we will see, a crucial ingredient is the light-like cusp anomalous dimension
$\gamma_{K, r} (\alpha_s)$, governing the UV divergences of form factors
of Wilson lines in representation $r$~\cite{Dotsenko:1979wb,Brandt:1981kf,
Korchemsky:1985xj,Korchemsky:1985xu,Korchemsky:1987wg}. For the sake 
of simplicity, we will consider the approximation in which the cusp anomalous 
dimension satisfies `Casimir scaling', {\it i.e.} we will write
\beq
  \gamma_{K, r} (\alpha_s) \, = \, C_r^{(2)} \, \widehat{\gamma}_K (\alpha_s) \, ,
\label{Cascal}
\eeq
where $\widehat{\gamma}_K (\alpha_s)$ is representation-independent. This
approximation is known to fail at the four-loop level~\cite{Moch:2018wjh,Grozin:2017css,
Henn:2019rmi,Henn:2019swt,vonManteuffel:2020vjv,Falcioni:2020lvv}, where 
contributions proportional to quartic Casimir eigenvalues have been shown to 
appear, with expected consequences on the structure of $\Gamma_n$ at four 
loops and beyond~\cite{Ahrens:2012qz,Becher:2019avh}. With this single 
simplifying approximation, the soft anomalous dimension matrix admits the 
all-order representation
\beq
  \Gamma_n \bigg( \frac{p_i}{\mu}, \alpha_s(\mu) \bigg) \, = \, 
  \Gamma_n^{\rm \, dipole} \bigg( \frac{s_{ij}}{\mu^2}, \alpha_s(\mu) \bigg) +
  \Delta_n \Big( \rho_{ijkl}, \alpha_s (\mu) \Big) \, .
\label{AllGamma}
\eeq
The colour-dipole term $\Gamma_n^{\rm \, dipole}$ is the only contribution at one
and two loops, and can be written as
\beq
  \Gamma_n^{\rm \, dipole} \bigg( \frac{s_{ij}}{\mu^2}, \alpha_s(\mu) \bigg) \, = \,
  \frac{1}{2} \, \widehat{\gamma}_K \big( \alpha_s(\mu) \big) \sum_{i = 1}^n 
  \sum_{j = i + 1}^n \log  \frac{- s_{ij} + {\rm i} \eta}{\mu^2} \,\, {\bf T}_i \cdot {\bf T}_j 
  \, - \, \sum_{i = 1}^n \gamma_i \big( \alpha_s (\mu) \big) \, ,
\label{GammaDip}
\eeq
where we assumed for simplicity that all particles are outgoing, $s_{ij} = 2 p_i 
\cdot p_j$, and $\gamma_i (\alpha_s)$ is the UV anomalous dimensions for the 
field corresponding to particle $i$. The conformal correction $\Delta_n$ arises
starting at three loops, with at least four particles, and it is constrained by
scale invariance to depend only upon the conformal cross-ratios
\beq
  \rho_{ijkl} \, = \, \frac{p_i \cdot p_j \, p_k \cdot p_l}{p_i \cdot p_l \, p_j \cdot p_k} 
  \, = \, \frac{s_{ij} s_{kl}}{s_{il} s_{jk}} \, .
\label{cicr}
\eeq
At the three-loop level, $\Delta_n$ was computed in Refs.~\cite{Almelid:2015jia,
Henn:2016jdu,Almelid:2017qju}, and it is built out of quadrupole correlations 
of the form
\beq
  F_{ijkl} \big( \{ \rho \} \big) \, f_{abe} f_{cd}^{\,\,\,\,\, e} \, {\bf T}_i^a
   {\bf T}_j^a {\bf T}_k^c {\bf T}_l^d \, ,
\label{StruDel}
\eeq
where the kinematic dependence is contained in the functions $F_{ijkl}$; by Bose 
symmetry and colour conservation, actually only one function is involved, and it
turns out to be an extremely simple combination of weight-5 single-valued harmonic
polylogarithms~\cite{Brown:2004ugm,Schnetz:2013hqa}.

In the present context, it is important to note that the dipole term can be organised
in a more useful way by exploiting colour conservation to simplify the scale dependence,
which will need to be integrated. One may write
\beq
  \Gamma_n^{\rm \, dipole} \left( \frac{s_{ij}}{\lambda^2}, \alpha_s (\lambda, \e) \right) 
  & = & \frac{1}{2} \, \widehat{\gamma}_K \big(\alpha_s (\lambda, \e) \big)
  \sum_{i = 1}^n \sum_{j = i + 1}^n \ln \left( \frac{- s_{ij} + {\rm i} \eta}{\mu^2} \right)
  {\bf T}_i \cdot {\bf T}_j \nonumber \\
  & & - \, \sum_{i = 1}^n \gamma_i \big(\alpha_s (\lambda, 
  \e) \big) 
  \, - \, \frac{1}{4} \, \widehat{\gamma}_K \big(\alpha_s (\lambda, 
  \e) \big) \ln \left(\frac{\mu^2}{\lambda^2}\right)
  \sum_{i = 1}^n C_i^{(2)} \nonumber \\
  & \equiv & \Gamma_n^{\rm \, corr.} \bigg( \frac{s_{ij}}{\mu^2}, 
  \alpha_s (\lambda, \e) \bigg) 
  \, + \, \Gamma_n^{\rm \, singl.} \bigg( \frac{\mu^2}{\lambda^2}, 
  \alpha_s (\lambda, \e) \bigg) \, ,
\label{Gammasimp}
\eeq
where $\mu$ is a fixed scale, while $\lambda$ is the integration variable in \eq{solZ}.
In this form, one can show that (aside from poles originating from the running of 
the coupling) the first term, which contains dipole colour correlations, generates 
only single poles of soft origin, while the remaining terms, which are colour singlets, 
generate single poles of hard-collinear origin, as well as double poles of soft-collinear 
origin. In order to compute them explicitly, one may use the explicit expression for the 
$d$-dimensional running coupling at the desired order, and perform the scale integration 
in \eq{solZ}. For example, at the one-loop level, one may use the lowest-order expression
for $d$-dimensional running coupling
\beq
  \alpha_s \left( \lambda, \e \right) \, = \, \alpha_s \left( \mu \right) 
  \left( \frac{\lambda^2}{\mu^2} \right)^{\! - \e} \, ,
\label{treealph}
\eeq
together with the basic integrals
\beq
  \int_0^{\mu^2} \frac{d \lambda^2}{\lambda^2} \, \alpha_s \!
  \big( \lambda, \e \big) \, = \, - \, \frac{1}{\eps} \, \alpha_s (\mu) \, , \quad
  \int_0^{\mu^2} \frac{d \lambda^2}{\lambda^2} 
  \ln \left( \frac{\lambda^2}{\mu^2} \right) \alpha_s \!
  \big( \lambda, \e \big) \, = \,  - \, \frac{1}{\eps^2} \, \alpha_s (\mu) \, ,
  \quad \big( \e < 0 \big) \, .
\label{fundintans}
\eeq
Note that in a gauge theory which is conformal in $d = 4$, such as ${\cal N} = 4$
Super-Yang-Mills theory, the result in \eq{treealph} is exact, and indeed the 
logarithm of the infrared operator ${\cal Z}_n$ has only single and double 
poles~\cite{Bern:2005iz}, which are trivially determined in terms of the perturbative 
coefficients of the singlet anomalous dimensions $\widehat{\gamma}_K$ and 
$\gamma_i$. In what follows, we will concentrate on the colour correlated term, 
$\Gamma_n^{\rm \, corr.}$. It is useful to remember that that it originates 
from the Wilson-line correlator
\beq
  {\cal S}_n \Big( \beta_i \cdot \beta_j, \alpha_s (\mu), \e \Big) \, \equiv \,
  \bra 0 | \prod_{i = 1}^n \Phi_{\beta_i} (\infty, 0) \, | 0 \ket \, ,
\label{WilCor}
\eeq
where $\beta_i$ are dimensionless four-velocity vectors parallel to the momenta $p_i$,
and $\Phi_{\beta_i}$ are semi-infinite Wilson lines along the classical trajectories
of the particles exiting the hard interaction vertex
\beq
  \Phi_{\beta} (\infty, 0) \, \equiv \, P \exp \left[ {\rm i} g \int_0^\infty d \lambda
  \beta \cdot A (\lambda \beta) \right] \, .
\label{Willi}
\eeq
Note that the Wilson lines in \eq{WilCor} have open colour indices at both ends, 
so that ${\cal S}_n$, as announced, is a colour operator acting on the hard factor 
${\cal H}_n$ in \eq{factampl}. Classically, the correlator in \eq{WilCor} is invariant 
under the rescalings $\beta_i \to \kappa_i \beta_i$, however this invariance is 
broken by collinear divergences in the massless case: the dipole contribution 
to the anomalous dimension matrix arises as a solution to the corresponding 
`anomaly equation', where the `anomaly' is expressed by the cusp anomalous 
dimension~\cite{Gardi:2009qi,Becher:2009qa}.


\section{Colour correlations on the celestial sphere}
\label{sphecol}

Taking inspiration from Refs.~\cite{Almelid:2017qju,Kalyanapuram:2020epb}, 
we now rewrite the colour-correlated part of the infrared anomalous dimension 
matrix in terms of coordinates on the celestial sphere. We parametrise the momenta
$p_i$ as
\beq
  p_i^\mu \, = \, \omega_i \, \Big\{ 1 + z_i \zbar_i, \, z_i + \zbar_i, \, 
  - {\rm i} \big( z_i - \zbar_i \big), \, 1 - z_i \zbar_i \Big\} \, ,
\label{spherepar}
\eeq
which is real if (as we assume) $\zbar_i = z_i^*$, and implies
\beq
  s_{ij} \, = \, 2 p_i \cdot p_j \, = \, 4 \omega_i \omega_j \big| z_i - z_j \big|^2 \, ,
\label{sij}
\eeq
so that all momenta are indeed light-like, $p_i^2 = 0$. We can define the four-velocities
$\beta_i$, for example, by $p_i^\mu = \sqrt{2} \omega_i \beta_i^\mu$, so that $\beta_i \cdot 
\beta_j = |z_i - z_j|^2$. Then we notice that (as expected from \eq{WilCor}) the energies 
$w_i$ do not contribute to colour correlations, again as an effect of colour conservation.
Indeed one can use
\beq
  \log \big( \! - \! s_{ij} + {\rm i} \eta \big) \, = \, \log \Big( \big| z_i - z_j \big|^2 \Big) + 
  \log \omega_i + \log \omega_j +  2 \log 2 + {\rm i} \pi \, ,
\label{energies}
\eeq
and perform the colour sum on terms that do not depend simultaneously on $i$ and 
$j$. All these terms can then be shuffled into the colour-singlet contribution to $\Gamma_n$,
so that we can write
\beq
  \Gamma_n^{\rm \, dipole} \left( \frac{s_{ij}}{\lambda^2}, \alpha_s (\lambda, \e) \right) 
  \, \equiv \,
  \widehat{\Gamma}_n^{\rm \, corr.} \Big( z_{i j}, \alpha_s (\lambda, \e) \Big) 
  \, + \, \widehat{\Gamma}_n^{\rm \, singl.} \bigg( \frac{\omega_i}{\lambda}, 
  \alpha_s (\lambda, \e) \bigg) \, ,
\label{alGammasimp}
\eeq
where we defined $z_{ij} \equiv z_i - z_j$.  The colour-singlet contribution is given by
\beq
  \widehat{\Gamma}_n^{\rm \, singl.} \bigg( \frac{\omega_i}{\lambda}, 
  \alpha_s (\lambda, \e) \bigg) \, = \, 
  - \, \sum_{i = 1}^n \gamma_i \big(\alpha_s (\lambda, \e) \big) \,
  \, - \, \frac{1}{4} \, \widehat{\gamma}_K \big(\alpha_s (\lambda, 
  \e) \big) \sum_{i = 1}^n \,
  \ln \left( \frac{- 4 \hspace{1pt}  \omega_i^2 + {\rm i} \eta}{\lambda^2} \right)
  C_i^{(2)} \, ,
\label{FinSingl}
\eeq
while
\beq
  \widehat{\Gamma}_n^{\rm \, corr.} \Big( z_{ij}, \alpha_s (\lambda, \e) \Big) 
  \, = \, 
  \frac{1}{2} \, \widehat{\gamma}_K \big(\alpha_s (\lambda, \e) \big)
  \sum_{i = 1}^n \sum_{j = i + 1}^n \ln \Big( \big| z_{ij} |^2 \Big) \, 
  {\bf T}_i \cdot {\bf T}_j \, .
\label{FinCorr}
\eeq
The remarkable fact about \eq{FinCorr} is that the scale dependence is factorised, 
universal, and free from colour correlations. This enables us to write the 
colour-correlated part of the infrared operator ${\cal Z}_n$ in a strikingly 
simple way, as
\beq
  {\cal Z}_n^{\rm \, corr.} \Big( z_{ij}, \alpha_s(\mu), \e \Big) & \equiv & 
  \exp \left[ \int_0^{\mu} \frac{d \lambda}{\lambda} \,\, 
  \widehat{\Gamma}_n^{\rm \, corr.} \Big( z_{ij}, \alpha_s (\lambda, \e) \Big) \right] 
  \nonumber \\
  & = & 
  \exp \bigg[ - K \big( \alpha_s (\mu), \e \big) \, \sum_{i = 1}^n \sum_{j = i + 1}^n 
  \ln \Big( \big| z_{ij} |^2 \Big) \, {\bf T}_i \cdot {\bf T}_j \bigg] \, ,
\label{ZCorr}
\eeq
where the universal scale-dependent prefactor is given by
\beq
  K \big( \alpha_s (\mu), \e \big) \, = \, - \frac{1}{2} \int_0^\mu \frac{d \lambda}{\lambda}
  \,\, \widehat{\gamma}_K \big( \alpha_s( \lambda, \e) \big) \, .
\label{Kdef}
\eeq
The function $K$ is well-known in QCD, where it plays a role in a number of different
settings: importantly, it defines the perturbative Regge trajectory in the high-energy
limit of $2 \to 2$ scattering amplitudes~\cite{Korchemsky:1993hr,Korchemskaya:1994qp,
Korchemskaya:1996je,DelDuca:2011xm,DelDuca:2011ae}, and it computes soft-collinear 
poles for Sudakov form factors~\cite{Magnea:1990zb,Magnea:2000ss}. In a gauge 
theory with conformal symmetry in $d = 4$ one finds
\beq
  K \big( \alpha_s, \e \big) \, = \, \sum_{n = 1}^\infty 
  \bigg( \frac{\alpha_s}{\pi} \bigg)^{\! n} \, \frac{\widehat{\gamma}_K^{(n)}}{4 n \e} \, ,
\label{Kconf}
\eeq
where $\widehat{\gamma}_K^{(n)}$ are the coefficients of the expansion of the
function $\widehat{\gamma}_K (\alpha_s)$ in powers of $\alpha_s/\pi$. With a 
non-vanishing $\beta$ function one finds instead~\cite{DelDuca:2014cya}
\beq
  K \big( \alpha_s, \e \big) & = & \frac{\alpha_s}{\pi} 
  \frac{\widehat{\gamma}_K^{(1)}}{4 \e} + \bigg( \frac{\alpha_s}{\pi} \bigg)^{\! 2}
  \Bigg( \frac{\widehat{\gamma}_K^{(2)}}{8 \e} + 
  \frac{b_0 \widehat{\gamma}_K^{(1)}}{32 \e^2} \Bigg) 
  \nonumber \\ && \hspace{1.2cm} + \, 
  \bigg( \frac{\alpha_s}{\pi} \bigg)^{\! 3} 
  \Bigg( \frac{\widehat{\gamma}_K^{(3)}}{12 \e} + 
  \frac{b_0 \widehat{\gamma}_K^{(2)} + b_1 \widehat{\gamma}_K^{(1)}}{48 \e^2} 
  + \frac{b_0^2 \, \widehat{\gamma}_K^{(1)}}{192 \e^3} \Bigg)
  + {\cal O} \Big( \alpha_s^4 \Big) \, ,
\label{Kgen}
\eeq
and all the higher-order coefficients can be explicitly determined~\cite{Magnea:2000ss}
in terms of $b_n$ and $ \widehat{\gamma}_K^{(n)}$. A simple factorised form for
the infrared operator ${\cal Z}_n$ in terms of the function $K$ had previously been
observed in the high-energy limit~\cite{Korchemskaya:1994qp,Korchemskaya:1996je,
DelDuca:2011xm,DelDuca:2011ae}, but the complete generality of \eq{ZCorr} only
becomes apparent in the celestial coordinates of~\eq{spherepar}.

As noted in Ref.~\cite{Kalyanapuram:2020epb} in the abelian case (where the 
operators ${\bf T}_i$ are replaced by the electric charges $e_i$, and the one-loop
approximation to the cusp anomalous dimension suffices), \eq{ZCorr} bears a 
striking resemblance to a correlator of vertex operators in a $d=2$ conformal 
theory. In the non-abelian case, we need to handle the matrix structure of 
\eq{ZCorr}, but the eventual connection to conformal invariance is actually 
strengthened by the knowledge that high-order corrections depend only on 
conformal cross-ratios, which in this language read $\rho_{ijkl} = | z_{ij} |^2 
| z_{kl} |^2/\big(| z_{il} |^2 | z_{kj} |^2 \big)$. In what follows, we will make a 
proposal for the conformal theory generating \eq{ZCorr}, and speculate about 
the origin of higher-order corrections.


\section{A theory of Lie-algebra-valued free bosons}
\label{matfrebo}

\noindent
We consider a set of scalar fields on a two-dimensional sphere, $\phi_a (z, \zbar)$, 
forming a multiplet in the adjoint representation of a Lie algebra, which we will take 
to be $su(N_c)$, so that $a = 1, \ldots, N_c^2 - 1$. For these fields one can naturally
choose the action
\beq
  S (\phi) \, = \,  \frac{1}{2 \pi} \int d^2 z \, \partial_z \phi^a (z, \zbar) \, 
  \barpartial \phi_a (z, \zbar) \, ,
\label{action}
\eeq
so that we have a theory of Lie-algebra-valued free bosons. Since we will have to
pair these free bosons with colour operators, one may also note from the outset 
that the scalars could be naturally organised into a matrix field
\beq
  \Phi_r (z, \zbar) \, \equiv \, \phi_a (z, \zbar) \, T^a_{r, \, z} \, ,
\label{matrix}
\eeq
where $T^a_{r, \, z}$ are generators of the group in the irreducible representation 
$r$. Considering the gauge-theory structure we are trying to mimick, the generators 
must be associated with the point $z$ on the sphere where the fields are evaluated: 
in particular, this means that the colour-space commutator $\big[ \Phi (z_1, \zbar_1), 
\Phi (z_2, \zbar_2) \big]$ must vanish, so long as $z_1 \neq z_2$. Products of matrix 
fields at different points are operators on the tensor product of the corresponding 
irreducible representations, and indeed one is free to consider products of matrix 
fields belonging to different representations. In terms of these matrix fields, one
can write the action in \eq{action} as
\beq
  S (\Phi) & = & \frac{1}{2 \pi t_r} \int d^2 z \, {\rm Tr} \Big[ \partial_z \Phi_r (z, \zbar) \, 
  \barpartial  \, \Phi_r (z, \zbar) \Big] \, ,
\label{alaction}
\eeq
where we normalised the generators according to ${\rm Tr} \left( T_r^a T_r^b \right) 
\, = \, t_r \delta^{ab}$. Writing the action in the form of \eq{alaction}, with matrices 
$T^a_{r, \, z}$ that are taken to depend on $z$, suggests that the proper underlying 
mathematical structure should be a gauge bundle on the sphere, and one might then 
expect that the derivative should be replaced by a covariant derivative. Such an 
extension is non-trivial for at least two reasons: first of all, one would need a precise 
form for the gauge connection on the celestial sphere; next, we emphasise that 
the collinear limits where two punctures $z_i$ and $z_j$ coincide is singular in
our framework: indeed, \eq{factampl}, which was our starting point, was derived
for {\it fixed-angle} scattering amplitudes, where none of the invariants $s_{ij}$
vanishes. In any case, we expect that corrections to \eq{alaction} arising from 
this issue would be of higher order in the gauge coupling, and furthermore they would 
contribute only to colour structures beyond dipoles, involving the structure constants 
$f_{abc}$. The free action in \eq{action} will therefore suffice for our present 
purposes, and we will take it as the definition of the theory. We will briefly
come back to this issue in \secn{Open}.

It is clear that in this setup the nature of the fields $\phi^a (z, \zbar)$ as fields in a 
conformal theory on the sphere is quite decoupled from the matrix structure that we 
have superimposed. Indeed, \eq{action} is essentially the bosonic string action in a 
conformal gauge, with scalar fields interpreted as coordinates in a Lie algebra 
rather than a spacetime -- in a sense, it is a colour-kinematic dual~\cite{Bern:2008qj} 
of the tree-level bosonic string. Most of the well-known formalism for the treatment 
of two-dimensional free scalar fields~\cite{Ginsparg:1988ui} is then simply inherited 
by our theory with only minor changes. In particular, the equations of motion are
\beq
  \partial_z \, \barpartial \, \phi^a (z, \zbar) \, = \, 0 \, ,
\label{EoM}
\eeq
implying that the field $\partial_z \phi^a$ is holomorphic, while $\barpartial \phi^a$
is anti-holomorphic. In the quantum theory, when taking matrix elements of products
of fields, \eq{EoM} leads to
\beq
  \partial_z \, \barpartial \, \phi^a (z, \zbar) \, \phi^b (w, \overline{w}) \, = \, 
  - \, \pi \, \delta^{ab} \delta^2 (z - w, \zbar - \overline{w}) \, ,
\label{EoMQ}
\eeq
which suggests the definition of the normal-ordered product
\beq
  : \! \phi^a (z, \zbar) \, \phi^b (w, \overline{w}) \! : \, \, = \, 
  \phi^a (z, \zbar) \, \phi^b (w, \overline{w}) + \frac{1}{2} \,
  \delta^{ab} \log \left| z - w \right|^2 \, ,
\label{normord}
\eeq
whose derivatives are free of contact terms. Following standard arguments, the 
theory defined by \eq{action} has a traceless conserved energy momentum tensor, 
which can be organised into holomorphic and anti-holomorphic components, with
\beq
  T(z) \, = \, - \, : \! \partial_z \phi^a (z, \zbar) \, \partial_z \phi_a (z, \zbar) \! : \, ,
  \qquad
  \widetilde{T}(\zbar) \, = \, - \, : \! \barpartial \phi^a (z, \zbar) \, 
  \barpartial \phi_a (z, \zbar) \! : \, .
\label{enmomt}
\eeq
More interestingly, since the action in \eq{action} involves only derivatives of 
$\phi^a$, there is a symmetry under translations in field space, which are now 
to be interpreted as translations in the Lie algebra. The Noether current for this
symmetry is Lie-algebra valued, and we will propose that it should be interpreted 
as a leading-order contribution to a Kac-Moody current for the full theory, to be 
identified with the current constructed in~\cite{Strominger:2013lka,He:2015zea}. 
Its holomorphic and anti-holomorphic components are simply given by the 
derivatives of the scalar fields
\beq
  j^a (z) \, = \, \partial_z \phi^a (z, \zbar) \, , \qquad 
  \tilde{j}^a (\zbar) \, = \, \partial_{\zbar} \phi^a (z, \zbar) \, .
\label{maykac}
\eeq
Following the standard lore for free bosons, we may now introduce vertex operators 
for this theory, in the form
\beq
  V (z, \zbar) \, \equiv \, : \, {\rm e}^{{\rm i} \kappa \, 
  {\bf T}_{\! z} \cdot \, {\bf \phi} (z, \zbar)} :
  \,\, = \,\, : \, {\rm e}^{{\rm i} \kappa \Phi (z, \zbar)} : \, ,
\label{vertex}
\eeq
where we dropped the representation label $r$, and we introduced the
operator notation discussed in \secn{nonabsad}. The vertex operators in
\eq{vertex} are matrices in the selected representation of the gauge group:
one can think of them as operators defined on the celestial sphere and
acting on the bulk colour degrees of freedom. We have introduced a 
coupling $\kappa$ in the exponent, allowing for the fact that the normalisation 
of the fields at this stage is arbitrary, and we will need to match our results to the 
gauge theory. The vertex operators in \eq{vertex} are reminiscent of those used in 
the vertex-operator construction of Kac-Moody algebras~\cite{Goddard:1986bp},
but we emphasise that they are quite different. In the Kac-Moody construction,
one works in the Cartan-Weyl basis for the Lie algebra, and one introduces
a (Fubini-Veneziano) scalar field $Q_i (z, \zbar)$ only for generators $H_i$ in 
the Cartan subalgebra; one then builds a vertex operator for every root 
$\alpha$ of the algebra, with an exponent proportional to $\alpha^i Q_i$.
These operators build a representation of the Kac-Moody algebra in the
Hilbert space of the conformal theory, but they act as numbers in colour 
space. In our case, on the other hand, in order to reproduce the gauge 
theory results, we need an expression treating all colour degrees of freedom 
on the same footing, which results in colour matrices. Pursuing an alternative 
analogy, one may note that a standard construction of Kac-Moody algebras 
from ordinary Lie algebras involves promoting the parameters $\theta^a$ of 
the Lie algebra to functions $\theta^a(z)$ on a circle $S^1$, and taking Fourier 
modes: in a sense, \eq{vertex} is similar, with the circle replaced by the sphere 
$S^2$. Finally, we must note a formal analogy between the fields $\phi^a$ in 
\eq{vertex} and the interpolating fields for Reggeized gluons introduced 
in Ref.~\cite{Caron-Huot:2013fea}, where however the gauge-theory Wilson 
lines are not semi-infinite but infinite, and the sphere is replaced by the 
transverse plane. For the purposes of the present paper, we will simply 
take \eq{vertex} at face value, and derive its properties.

With the operators in \eq{vertex}, we can construct correlation functions 
which bear a direct analogy with the world-sheet integrands of tree-level 
bosonic string amplitudes. There are a number of potential obstacles for the
consistency of this procedure, that we will now consider, but, interestingly, 
they can all be bypassed using well-understood properties of the operators 
${\bf T}_{\! z}$.
\begin{itemize}
\item First of all, we need to establish that the vertex operator in \eq{vertex}
is a good conformal field. This appears far from obvious given its matrix 
structure. Once again, following standard procedures, one can write down 
an explicit solution of the equations of motions for the fields $\phi_a(z, \zbar)$
in terms of a Fourier expansion, and proceed to quantise the theory by interpreting 
the coefficients of the Fourier modes as (rescaled) harmonic-oscillator creation 
and annihilation operators. In this formalism, the conformal dimension 
$h \, (= \bar{h})$ of the vertex operator emerges entirely from the normal 
ordering, bringing annihilation operators to the left and creation operators 
to the right. The colour operators ${\bf T}_{\! z}$ are spectators in this 
calculation, and the standard free-field (`bosonic string') result is reproduced. 
Pursuing the analogy, we recall that in a string-theory context one finds
\beq
  V_{\rm c.s.} (z, \zbar) \, \equiv \, : \, {\rm e}^{{\rm i} k^\mu X_\mu (z, \zbar)} :
  \qquad \longrightarrow \qquad h \, = \, \frac{1}{4} \, k^\mu k^\nu \eta_{\mu \nu} 
  \, = \, \frac{k^2}{4} \, ,
\label{bostrivertex}
\eeq
where the space-time metric $\eta_{\mu \nu}$ comes from the interpretation
of the fields $X^\mu$ as space-time coordinates. In the case of \eq{vertex},
the fields $\phi_a$ are Lie-algebra coordinates, and the metric is simply 
the Cartan-Killing metric $\delta_{ab}$. Therefore
\beq
  V (z, \zbar) \, \equiv \, : \, {\rm e}^{{\rm i} \kappa \, 
  {\bf T}_{\! z} \cdot \, {\bf \phi} (z, \zbar)} :
  \qquad \longrightarrow \qquad h \, = \, \frac{\kappa^2}{4} \, {\bf T}_{\! z}
  \cdot {\bf T}_{\! z} \, = \, \frac{\kappa^2}{4} \, C_r^{(2)} \, ,
\label{confdimvertex}
\eeq
where $C_r^{(2)}$ is the quadratic Casimir eigenvalue for representation $r$, 
which, crucially, is a number. In a conformal theory, this result must be consistent
with the scaling properties of two-point functions of vertex operators, which
must satisfy
\beq
  \big \langle V (z_1, \zbar_1) V (z_2, \zbar_2) \big \rangle \, \sim \, 
  | z_{12} |^{- 2 \Delta} \, ,
\label{twopf}
\eeq
with $\Delta = h + \bar{h}$. As we will see below, this is indeed the case, upon
enforcing colour conservation, which for the two-point function requires ${\bf T}_1
= - {\bf T}_2$.
\item Of course, the analogy with strings should not be pushed too far: for example, 
we note that, in the case of a string theory, the value of $k^2$ in \eq{bostrivertex} is 
fixed, and allowed values for different vertex operators give the mass spectrum 
of the string. The origin of the constraint, however, is the fact that string-theory 
vertex operators relevant to string amplitudes are {\it integrated} over the world 
sheet, to preserve diffeomorphism invariance. Invariance under rescalings of the 
world-sheet coordinates then imposes $h = 1$ for the (`tachyon') vertex operator 
we are considering, and integer values of $h$ for vertex operators involving prefactors 
with derivatives of the fields. In our case, we will never need to integrate over the 
locations of the punctures, which represent the (fixed) momenta of the hard particles, 
so there are no constraints of this kind on $h$. We are genuinely interested in the
conformal correlation function in a fixed coordinate system.
\item We will be evaluating correlation functions of $n$ vertex operators of the form of 
\eq{vertex}. In the case of bosonic strings, arguments from holomorphicity, or just the 
evaluation of such correlators from the path integral on the sphere~\cite{Polchinski:1998rq}, 
show that the results are consistent and scale-invariant only if momentum is conserved, 
$\sum_i k_i^\mu = 0$. For string amplitudes, this has an obvious physical interpretation. 
In the present case, colour conservation takes the place of momentum conservation, as 
noted in \secn{nonabsad}: the condition $\sum_i {\bf T}_{\! z_i} = 0$, which must be 
satisfied when the correlator is regarded as an operator acting on the `bulk' gauge 
theory, guarantees the proper scale-invariant behaviour. Correlators will be operators 
with support only on colour-conserving quantities, consistently with the gauge-invariance 
of the four-dimensional theory.
\end{itemize}

\noindent
Having established that we have a bona fide conformal field theory on the sphere, 
and well-behaved -- though matrix-valued -- vertex operators, the obvious next 
step is to study the correlator
\beq
  {\cal C}_n \Big( \{ z_i \}, \kappa \Big) \, \equiv \, 
  \Big \langle \prod_{i = 1}^n V (z_i, \zbar_i) \Big \rangle \, .
\label{corre}
\eeq
The evaluation of the correlator ${\cal C}_n$ is a textbook exercise (see for example 
Chapter 6 in Ref.~\cite{Polchinski:1998rq}): in a conformal invariant theory,
one is always allowed to use a locally flat metric (as we are doing), and one simply
gets an expression analogous to the integrand of the Virasoro-Shapiro amplitude 
for tree-level closed strings. We find
\beq
  {\cal C}_n \Big( \{ z_i \}, \kappa \Big) \, = \, C(N_c) \,
  \exp \left[ \frac{\kappa^2}{2} \sum_{i = 1}^n \sum_{j = i + 1}^n 
  \ln \Big( \big| z_{ij} |^2 \Big) \, {\bf T}_i \cdot {\bf T}_j  \right] \, ,
\label{riscorr}
\eeq
where $C(N_c)$ is a constant dependent on the Lie algebra. This is of course the
result we have been working to achieve: it is precisely of the form of \eq{ZCorr}, which 
in turn generates {\it all} infrared singularities with a colour-dipole structure for any 
massless gauge theory and to {\it any perturbative order}. It is therefore tempting 
at this stage to simply state that we can set the coupling $\kappa$ to reproduce 
\eq{ZCorr} exactly. This is certainly possible, but we note at the outset that this
identification raises interesting dynamical questions, concerning both the scale 
dependence of the answer, and dependence on the dimensional regulator $\e$.
We discuss this issue and some of the many important open questions in 
\secn{Open}. A further observation on \eq{riscorr} is warranted: if one computes 
the correlator in \eq{corre} by means of a path integral on a curved two-dimensional
surface, as done for example in Ref.~\cite{Polchinski:1998rq}, \eq{riscorr} is modified
by an extra factor arising from the Weyl anomaly. This factor cancels in the conformal
theory, but would be present, for example, if one were to break scale invariance
by considering a fixed sphere of finite radius $R$, thus introducing a length scale.
The Weyl factor for the correlator in \eq{corre} takes the form
\beq
  {\cal W}_n \Big( \{ z_i \}, \kappa \Big) \, = \, \exp \bigg[ \! - \frac{1}{2} \, \sum_{i = 1}^n
  C_i^{(2)} g(z_i, \zbar_i) \bigg] \, ,
\label{Weyl}
\eeq
where $g(z_i, \zbar_i)$ is a scale factor. The interesting point about \eq{Weyl} is
that the Weyl factor matches the factor arising in the gauge theory from the
energy-sensitive terms that we have shuffled to the colour-singlet part of the
anomalous dimension, $\widehat{\Gamma}_n^{\rm \, singl.}$ in \eq{FinSingl}.
Such `diagonal' contributions to the infrared operator ${\cal Z}_n$, with an exponent 
proportional to a sum of terms associated with single points $z_i$, cannot be
expected to emerge from the celestial CFT: on the gauge-theory side, they are 
related to collinear divergences, and do not contribute to colour correlations,
but they are sensitive to particle energies. 

There are further significant tests of the correspondence we have found between 
the infrared sector of non-abelian gauge theories and the proposed conformal 
theory on the celestial sphere, arising from soft and collinear limits. On the 
gauge-theory side, it is well-known that the tree-level emission of a soft gluon
factorises~\cite{Bassetto:1984ik} from hard $n$-point scattering amplitudes (and 
similarly from correlators of $n$ Wilson-lines such as ${\cal S}_n$ in \eq{WilCor}), 
and the soft factor is given by the soft current for the emission of a gluon of 
momentum $k$
\beq
  {\bf J}^\mu (k) \, = \, g \, \sum_{i = 1}^n {\bf T}_i \, 
  \frac{\beta_i^\mu}{\beta_i \cdot k} \, ,
\label{softglu}
\eeq
exposing the singular behaviour of the radiative amplitude as the gluon energy 
vanishes, as well as a set of collinear singularities as $k$ becomes parallel to 
$\beta_i$. The current is gauge-invariant, in the sense that its longitudinal component 
vanishes, since 
\beq
  k \cdot  {\bf J}^\mu (k) \, = \, g \, \sum_{i = 1}^n {\bf T}_i \, = \, 0 \, ,
\label{gaugeinv}
\eeq
by colour conservation. Remarkably, this factorisation theorem for soft non-abelian
radiation can be derived in the conformal theory on the sphere, by consider a 
generalised correlator of vertex operators, including an insertion of the Noether 
current in \eq{maykac}. Taking the OPE of the current with the vertex operators 
yields the result
\beq
   \Big \langle \partial_z \phi^a (z, \zbar) \, \prod_{i = 1}^n V (z_i, \zbar_i) 
   \Big \rangle \, \simeq \, - \frac{\rm i}{2} \sum_{i = 1}^{n} 
   \frac{{\bf T}_i^{\, a}}{z - z_i} \, {\cal C}_n \Big( \{ z_i \}, \kappa \Big) \, .
\label{softglucur}
\eeq
As discussed in detail in Ref.~\cite{He:2015zea} (see also Refs.~\cite{Strominger:2013lka,
Cheung:2016iub,Nande:2017dba,Fan:2019emx}), the {\it r.h.s.} of \eq{softglucur} 
is the representation of the soft-gluon theorem~\cite{Bassetto:1984ik} on the 
celestial sphere, and the poles as $z \to z_i$ are {\it collinear} poles. As noted 
above, in the gauge-theory this pole is superimposed to an angle-independent 
soft singularity arising when the energy of the soft gluon becomes vanishingly 
small; the soft singularity is absent in \eq{softglucur}, which can be understood 
in the present context by recalling that soft-collinear singularities are color-uncorrelated, 
as shown in \secn{nonabsad} for virtual corrections. Colour conservation emerges 
in \eq{softglucur} from the request that the Noether current be holomorphic as 
$z \to \infty$: the correlator in \eq{softglucur} must then vanish as $z^{-2}$ for 
large $z$, which indeed requires that $\sum_i {\bf T}_i = 0$. Importantly, 
Ref.~\cite{He:2015zea} finds the same result by employing a current constructed in 
terms of asymptotic expressions for the four-dimensional fields: this provides clear 
evidence that the two currents are different representations of the same object, at least 
to the accuracy of the present calculation. In the framework of Ref.~\cite{He:2015zea}, 
the soft pole is absent because the current is constructed out of field strengths rather 
than gauge potentials, providing an extra power of the energy in the numerator.
We note that one finds an equivalent result using the anti-holomorphic current
in \eq{maykac}: Ref~\cite{He:2015zea}, examining double soft emission, argues 
that one of the two currents generates a Kac-Moody symmetry, but there are
subtleties in attempting to extend this to both currents, related to the ordering
of soft limits in the double-emission matrix element. An analysis of this ambiguity
goes beyond the approximation we have taken so far (of including only 
colour-dipole correlations), since in QCD the difference between soft orderings
is proportional to colour correlations involving the structure constants $f_{abc}$,
but this is without doubt a very interesting topic for future work.

Next, we turn to collinear limits, when the momenta of two hard particles become
proportional, and consequently two of the punctures on the celestial sphere
become close. As we noted, strict collinear limits, where some of the Mandelstam
invariants $s_{ij}$ vanish, violate the factorisation theorem in \eq{factampl}, and
therefore lie outside the reach of our discussion\footnote{Tools more suitable to 
handle collinear limits in this context could possibly be devised by considering
directly the light-ray operators in the $d=4$ theory. These have a long history
in the context of QCD (see, for example, Refs.~\cite{Balitsky:1987bk,Braun:2003rp}),
and have been intensively studied recently in a CFT context (see, for example,
Refs.~\cite{Kravchuk:2018htv,Kologlu:2019mfz}). Interestingly, phenomenological
applications of these tools to QCD have already stated to appear~\cite{Chen:2020adz,
Chen:2021gdk}.}. We can, however, study the approach to the collinear limit,
when two particles $i$ and $j$ are nearly aligned, so that the invariant $s_{ij}$
is much smaller than all other invariants. In this limit, the factorisation in \eq{factampl}
correctly captures all logarithms of the small invariant that are multiplied times
infrared poles. With this understanding, on the gauge theory side, collinear limits 
of \eq{GammaDip} are well understood, and tested up to three loops (where 
of course the correction $\Delta_n$ must also be taken into account). On the 
celestial side of the correspondence, collinear limits must correspond to short 
distance limits, $|z_{ij}| \to 0$, and therefore can be probed by considering
the OPE of the vertex operators in \eq{vertex}. In order to clarify the correspondence, 
let us begin by giving a precise definition of the collinear limit (see also 
Ref.~\cite{Ebert:2020nqf}). Following standard practices in QCD, in order 
to describe the situation in which two light-like momenta $p_1$ and
$p_2$ become collinear along a light-like direction $p$, we introduce
the Sudakov parametrisation
\beq
  p_1^\mu & = & x p^\mu + p_\perp^\mu - 
  \frac{p_\perp^2}{2 x p \cdot n} \, n^\mu \, , \nonumber \\
  p_2^\mu & = & (1 - x) p^\mu - p_\perp^\mu - 
  \frac{p_\perp^2}{2 (1 - x) p \cdot n} \, n^\mu \, ,
\label{Sudakov}
\eeq  
where $n^\mu$ is a reference light-like vector satisying $n^2 = n \cdot 
p_\perp = 0$, and $p_\perp^\mu$ is a space-like vector orthogonal to the
collinear direction, $p \cdot p_\perp = 0$. The collinear limit is parametrised
by taking $p_\perp^2 \to 0$, and, in the limit, $p_1 + p_2 = p$. It is not
difficult to translate the Sudakov parametrisation to the celestial sphere:
using \eq{spherepar} for the momenta $p_1$ and $p_2$, and picking a
fixed vector $n^\mu$, for example, as
\beq
  n^\mu \, = \, \frac{1}{2} \Big\{ 1, 0, 0, - 1 \Big\} \quad \rightarrow \quad
  n \cdot p_i \, = \, \omega_i \, , \quad 
  n \cdot p = \omega_1 + \omega_2 \equiv \omega \, ,
\label{nmu}
\eeq
one may determine the celestial expression for $p^\mu$, with the result
\beq
  p^\mu \, = \, \omega \, \Big\{ 1 + z \zbar, \, z + \zbar, \, - {\rm i} (z - \zbar), \,
  1 - z \zbar \Big\} \, ,
\label{spherep}
\eeq
where $z = x z_1 + (1 - x) z_2$, and $\omega_1 = x \omega$, while 
$\omega_2 = (1 - x) \omega$. Furthermore, $p_\perp^\mu$ can be written as
\beq
  p_\perp^\mu & = & \omega x (1 - x) \, \Big\{ 
  (1 - 2 x) \big(z_1 \zbar_2 + \zbar_1 z_2) + 2 x z_1 \zbar_1 
  - 2 (1 - x) z_2 \zbar_2, \,
  \nonumber \\ && \hspace{3mm}
  z_{12} + \zbar_{12}, \,
  - {\rm i} \big( z_{12} - \zbar_{12} \big) , \, 
  - (1 - 2 x) \big(z_1 \zbar_2 + \zbar_1 z_2) - 2 x z_1 \zbar_1 
  + 2 (1 - x) z_2 \zbar_2 \Big\} \, .
\label{spherepperp}
\eeq
The transverse momentum vector $p_\perp^\mu$ is antisymmetric under
the exchange of the labels 1$\leftrightarrow$2 (which implies $x \leftrightarrow 
1 - x$), as expected from \eq{Sudakov}, and one easily verifies that it satisfies 
\beq
  p_\perp \cdot n \, = \, p_\perp \cdot p \, = \, 0 \, , \qquad \quad 
  p_\perp^2 \, = \, - x (1 - x) s_{12} \, = \, - 4 x (1 - x) \omega_1 \omega_2 
  \big| z_{12} \big|^2 \, ,
\label{pperpprop}
\eeq
providing a precise connection between the short distance limit on the sphere
and the collinear limit in the bulk.

Let us now use the conformal OPE to study the short distance limit of the 
correlator ${\cal C}_n$ as, say, $z_1 \to z_2$. When the locations of the two 
vertex operators are brought together one finds
\beq
  : \, {\rm e}^{{\rm i} \kappa \, {\bf T}_1 \cdot \, {\bf \phi} (z_1, \zbar_1)} : \, \, 
  : \, {\rm e}^{{\rm i} \kappa \, {\bf T}_2 \cdot \, {\bf \phi} (z_2, \zbar_2)} : \, \,
  \, \sim \, \big| z_{12} \big|^{\kappa^2 {\bf T}_1 \cdot {\bf T}_2} \,
  : \, {\rm e}^{{\rm i} \kappa \, \big( {\bf T}_1 + {\bf T}_2 \big) \cdot \, 
  {\bf \phi} (z, \zbar)} : \,\, ,
\label{OPE}
\eeq
where we placed the field on the {\it r.h.s.} of \eq{OPE} at the point $z = x z_1 +
(1 - x) z_2$ associated with the collinear direction $p^\mu$, and corrections 
are suppressed by powers of $z_{12}$. The $n$-point correlator ${\cal C}_n$ 
in this limit is thus re-expressed in terms of the $(n-1)$-point correlator 
${\cal C}_{n-1}$, where however point $z$ carries the sum of the colour 
operators of two merging punctures. This is precisely what one expects on the 
gauge theory side~\cite{Catani:2003vu,Becher:2009qa}, where this property
is a consequence of collinear factorisation. The quantity of interest is the 
{\it splitting anomalous dimension}, defined by 
\beq
  \Gamma_{\rm Sp.} \big( p_1, p_2 \big) \, \equiv \,
  \Gamma_n \big( p_1, p_2, \ldots, p_n \big) - 
  \Gamma_{n - 1} \big( p, p_3, \ldots, p_n \big) 
  \Big|_{{\bf T}_p \to {\bf T}_1 + {\bf T}_2} \, .
\label{GamSplit}
\eeq
The short distance limit on the celestial sphere, by means of \eq{OPE}, 
identifies the form of the correlator that one needs to compute in order to 
analyse the collinear limit in the gauge theory. To make the comparison 
precise, one needs to extract $\widehat{\Gamma}_n^{\rm corr.}$ from 
${\cal C}_n$, and the appropriate form of $\widehat{\Gamma}_{n-1}^{\rm 
corr.}$, with recombined colour factors, from ${\cal C}_{n-1}$. Next, one needs 
to reinstate the dependence on the energies of individual hard particles,
since the collinear limit is sensitive to how the energy of the parent 
particle is subdivided between the two particles forming the collinear pair.
At that point, the calculation mirrors exactly the gauge theory procedure:
the precise location of the point $z$ on the arc joining $z_1$ and $z_2$
is irrelevant at leading power in $p_\perp$, and the result arises entirely
from the scaling of particle energies, $\omega_1 = x \omega$ and
$\omega_2 = (1 - x) \omega$. One finds
\beq
  \Gamma_{\rm Sp.} \big( p_1, p_2 \big) & = & \frac{1}{2} \,
  \widehat{\gamma}_K ( \alpha_s ) \bigg[ \ln \bigg( 
  \frac{- s_{12} + {\rm i} \eta}{\mu^2} \bigg) {\bf T}_1 \cdot {\bf T}_2 
  - \ln x \,\, {\bf T}_1 \cdot \big( {\bf T}_1 + \, {\bf T}_2 \big)
  \nonumber \\ && \hspace{3cm}  
  - \, \ln (1- x) \,\, {\bf T}_2 \cdot \big( {\bf T}_1 + {\bf T}_2 \big)
  \bigg] \, ,
\label{GammaSplitFin}
\eeq
in agreement with Ref.~\cite{Becher:2009qa}. One may argue that the result
in \eq{GammaSplitFin} was to some extent implicit in our earlier identification
of the soft anomalous dimension matrix with the logarithm of the conformal 
correlator: we note however that the emerging colour structure is correctly
identified by the OPE on the celestial sphere. The result is highly non-trivial 
in the gauge-theory context: first of all, it embodies the collinear factorisation
theorem, stating that the collinear limit depends only on the particles participating 
in the splitting, and colour correlations with the remaining $n-2$ non-collinear 
particles vanish at leading power; furthermore, \eq{GammaSplitFin} is an all-order 
result, collecting all contributions to $\Gamma_{\rm sp.}$ arising from dipole 
colour correlations (interesting corrections do of course arise when $\Delta_n$ 
is included).


\section{Some open questions}
\label{Open}

It is clear that the construction that we have so far proposed represents only a
first step in the exploration of the emerging correspondence, and leaves many 
open questions. In this section, we begin to discuss some of these questions.

First of all, we note, as we did after \eq{FinCorr}, that the remarkable all-order
correspondence between the conformal correlator ${\cal C}_n$ on the
celestial sphere and the infrared operator ${\cal Z}_n^{\, \rm corr.}$ is made 
possible by the striking factorisation of the coupling and scale dependence
from angular and colour variables in \eq{ZCorr}. This raises the question of
the nature of the correspondence between the coupling $\kappa$ in the
conformal theory and the function $K (\alpha_s, \epsilon)$ in the gauge 
theory. This question has two aspects, one concerning the choice of the 
gauge coupling, and the other concerning the integrated scale dependence.
With regards to the coupling, we note that one can bypass the need for
a perturbative expansion of the function $\widehat{\gamma}_K (\alpha_s)$
by simply choosing a scheme in which $\widehat{\gamma}_K$ itself plays 
the role of the coupling. The idea that the universal cusp anomalous dimension
should serve as a fundamental gauge coupling, at least in the infrared,
is certainly not new, and it has been raised in several contexts (see,
for example, Refs.~\cite{Catani:1990rr,Erdogan:2011yc,Grozin:2015kna,
Banfi:2018mcq,Catani:2019rvy}), providing ample circumstantial evidence 
that such a choice is physically sensible. 

Having taken this option, one must still note that the celestial coupling $\kappa$ 
does not directly correspond to the new gauge coupling, $\widehat{\gamma}_K$, 
but rather to its `average' over scales, $K$ in \eq{Kdef}. We believe that this aspect 
of the correspondence is dynamically interesting, and deserves closer examination. 
In a sense, one could argue that {\it any} matching between the coupling $\kappa$ 
on the celestial sphere and the gauge theory {\it must}  share some of the 
characteristics of $K$. After all, correlation functions on the sphere are finite 
and scale-independent, while we are attempting to reproduce infrared divergences 
in a scale-dependent theory: regulator dependence and scale dependence must 
therefore reside in the matching coefficient. From the gauge theory side, it is quite 
surprising that such a matching is at all possible, since scale dependence is expected 
in general to be entangled with colour and kinematics. Even in the case of conformal 
gauge theories in $d=4$, the analysis of IR divergences requires breaking conformal 
symmetry by the introduction of an appropriate factorisation scale, much like 
what happens in the UV case: dimensional regularisation to $d > 4$ simply 
provides an elegant and computationally efficient way to perform this 
symmetry breaking. Indeed, as we noted, {\it all} IR divergences of scattering 
amplitudes, in any gauge theory, are generated by the scale integration, 
which can be taken to extend to the IR region when $d>4$, while all 
anomalous dimensions appearing in the scale integrand are finite. 
That being said, it would be very interesting and appealing if the scale 
integration could be made part of the correspondence, or, in other words, 
if one could understand the meaning of the scale integration from the point 
of view of the conformal theory on the celestial sphere. We note that, from 
the gauge theory point of view, the IR operator ${\cal Z}_n$ acts on a hard, 
finite coefficient function which is located at the common origin of the Wilson 
lines in \eq{WilCor}. Distance along the Wilson lines thus can be considered 
as a proxy for the inverse of the factorisation scale, with infrared scales 
($\mu \to 0$) located at large distances ($\lambda \to \infty$ in \eq{Willi}),
and ultraviolet scales located at short distances. The scale integration in 
\eq{ZCorr}, in this picture, could be understood as an integration over a
scale factor for the celestial sphere, as it is brought from infinite distance 
to distance $1/\mu$. At this stage, of course, these considerations are 
purely speculative, and they could only be made precise once a more detailed
correspondence is constructed.

A second set of considerations must be devoted to the known corrections
to \eq{ZCorr} arising in the gauge theory, {\it i.e.} the functions of conformal
cross ratios of kinematic invariants that arise starting at the three-loop level,
for the scattering of at least four hard particles, and build up the operator
$\Delta_n$. Clearly, these corrections cannot arise from the free-boson 
theory that we have just described: the action in \eq{action} must be 
complemented with interaction terms. The nature of these interaction 
terms is strongly constrained by our knowledge on the gauge theory side 
of the correspondence. We note, for example, that the three-point correlator 
in our free-boson theory is trivial, in the sense that it does not contain any 
colour correlations: indeed, all `scalar' products ${\bf T}_i \cdot {\bf T}_j$ 
reduce to Casimir eigenvalues, as a consequence of colour conservation, 
${\bf T}_1 + {\bf T}_2 + {\bf  T}_3 = 0$, which implies that ${\bf T}_1 \cdot 
{\bf T}_2 =  ( C^{(2)}_3 - C^{(2)}_1 - C^{(2)}_2 )/2$, and similarly for the 
other two cyclic permutations\footnote{We note in passing that these colour
identities guarantee the proper scaling behaviour of three-point functions
of vertex operators, consistent with \eq{twopf}.}. When more than three hard 
particles participate in the scattering, three-point correlations can arise, and 
one might expect terms of the form $f_{a b c} {\bf T}_1^a {\bf T}_2^b {\bf T}_3^c$ 
to appear in $\Gamma_n$: indeed, just such correlations do arise at two loops in 
the scattering of massive hard particles~\cite{Becher:2009kw,Ferroglia:2009ep,
Ferroglia:2009ii,Mitov:2009sv,Kidonakis:2009ev,Chien:2011wz}. No such 
corrections are possible, however, in the massless theory, essentially because 
conformal cross-ratios necessarily involve at least four points. The analysis 
of Refs.~\cite{Almelid:2015jia,Almelid:2017qju} furthermore shows that
three-point correlations do arise in $\Gamma_n$ for $n > 3$, but they 
are constrained to be independent of kinematics, and their colour structure 
has the peculiar form
\beq
  f_{a b e} f^e_{\,\,\, c d} \, \Big\{ {\bf T}_i^a, {\bf T}_i^d \Big\} \,
  {\bf T}_j^b {\bf T}_k^c \, ,
\label{threep}
\eeq
with $i \neq j \neq k$; in particular, correlations of this form are necessary 
to preserve collinear factorisation. The special form of three-point correlations 
is very relevant to possible generalisations of \eq{action}, since the bootstrap
principle for two-dimensional conformal theories holds that four-point functions 
of conformal fields are determined once the three-point functions are given. 
In the case at hand, of course, the precise form of the full correction to the 
four-point function at the three-loop level is also known, involving the first 
`quadrupole' correction to \eq{ZCorr}. Similar considerations apply to the
soft-gluon current in \eq{softglucur}: in gauge theories, the current has an
all-order definition in terms of matrix element of Wilson lines, with a gluon
state replacing the vacuum state~\cite{Magnea:2018ebr}; this matrix element
is known both at one~\cite{Catani:2000pi} and two loops~\cite{Badger:2004uk,
Duhr:2013msa,Li:2013lsa}, where it exhibits a non-trivial, purely non-abelian
colour structure. It is clear that this abundant and detailed knowledge on the 
gauge-theory side of the correspondence will provide powerful constraints
and checks for any attempt to construct the completion of our matrix-valued
free-boson proposal.


\section{Perspectives}
\label{Perspe}

The formulation of the infrared problem for gauge theories and gravity in terms of
asymptotic symmetries of the $S$-matrix, initiated in Refs.~\cite{Strominger:2013lka,
Strominger:2013jfa}, provides a novel and interesting viewpoint to investigate 
long-distance dynamics. One of the most intriguing aspects of this approach is
the emergence of a `holographic'~\cite{Cheung:2016iub} correspondence between 
certain dynamical properties of massless gauge theories in $d = 4$ Minkowski 
space and their asymptotic behaviour on the celestial sphere at large light-cone 
times. What emerges is a re-interpretation of the soft limit of gauge amplitudes 
in terms of a two-dimensional conformal theory defined on the sphere: many 
elements of this conformal theory can be constructed by taking appropriate 
limits of gauge-theory quantities, and infrared factorisation properties of low-order 
perturbative gauge amplitudes can be understood, and in fact generalised, 
in terms of conformal objects defined on the sphere.

In this paper, inspired by the QED analysis in Refs.~\cite{Kalyanapuram:2020epb}, 
we took a different viewpoint, starting from the {\it all-order} infrared factorisation 
properties of massless gauge theory amplitudes, and attempting to find a 
characterisation of the universal infrared operators emerging from the 
factorisation in terms of a celestial conformal field theory. For simplicity, 
we have focused on colour correlations of dipole form, which are the only 
ones appearing up to three loops in the massless theory: we emphasise 
however that our results apply to all perturbative orders within this set, 
and they include both planar and non-planar correlations. A first strong 
indication that the celestial sphere is an appropriate arena for the discussion 
of infrared divergences came from the specific form of the exponentiation 
of kinematic dependence in \eq{ZCorr}: indeed, using the parametrisation 
in \eq{spherepar} exposes a complete factorisation of coupling and scale 
dependence from kinematic and colour variables, that had not previously been 
noticed, except in the high-energy limit. The simple and transparent expression 
in \eq{ZCorr} provides a strong suggestion for the identification of the appropriate 
celestial conformal theory, generalising~\cite{Kalyanapuram:2020epb} to the 
non-abelian theory. We find that the all-order gauge-theory result in \eq{ZCorr} 
is reproduced by a celestial theory of {\it Lie-algebra-valued free bosons}, by 
computing the correlator of $n$ vertex operators which are gauge-group matrices 
in the representations of the hard particles participating in the scattering. We have 
seen that the matrix nature of the vertex operators does not prevent them 
from being interpreted as conformal fields of definite weight, and we have 
seen that their correlators have good conformal properties as a consequence 
of the gauge invariance of the bulk gauge theory. Like ${\cal Z}_n^{\rm \, corr.}$ 
in \eq{ZCorr}, the conformal correlator must be interpreted as a colour operator, 
defined on the celestial sphere, and acting on the bulk colour degrees of 
freedom, which can be thought of as being located at the origin.

Both the known gauge-theory results and the structure of our proposed 
celestial theory point to the incompleteness of of the picture drawn so far.
On the celestial side, the fact that we have a local matrix theory suggests
that a gauge connection on the sphere might need to be introduced, and
consequently the theory is expected to become interacting. While we 
leave this generalisation to future work, we have noted that known
multipole colour correlations arising in the gauge theory starting at
the three-loop order will pose stringent constraints on any extension
of \eq{action}. Clearly, the next very significant goal in this program
would be to compute the three-loop value of the operator $\Delta_n$,
first determined in Ref.~\cite{Almelid:2015jia}, from conformal field 
theory data. This could open the way to the exploration of even 
higher-order correlations with novel, and possibly simpler, techniques, 
not associated with gauge-theory Feynman diagrams. In particular,
it would be of great interest to understand, in the conformal context,
colour correlations associated with higher-order Casimir operators of
the gauge algebra, which have been neglected here, but are known
to be present starting at four loops~\cite{Grozin:2017css,Moch:2018wjh,
Henn:2019rmi,Henn:2019swt,vonManteuffel:2020vjv,Ahrens:2012qz,
Becher:2019avh}, and whose role on the gauge-theory side is not yet 
fully understood.

Aside from this first layer of generalisation, several other interesting 
questions and possible developments will need to be explored. First
of all, the role of collinear singularities should be clarified: the conformal
theory, as written, determines only colour {\it correlations}, which originate
exclusively from soft gluon exchanges at wide angles. Collinear poles
are `colour-singlet' quantities, and are associated to single points on the
sphere, and not to arcs. On the other hand, the relatively simple structure
of infrared singularities embodied by \eq{AllGamma} is specific to massless 
theories and thus it is inextricably linked to the presence of collinear poles.

A related, and certainly more complex question is to what extent this 
framework could be extended to massive hard particles: in a sense,
one would expect drastic complications, since the appropriate asymptotic
configuration space would no longer be the punctured Riemann sphere,
but rather a three-dimensional hyperbolic space; on the other hand, on
the gauge theory side, soft poles are still generated by correlators of 
semi-infinite Wilson lines, which are scale-invariant even away from 
the light-cone, and the modifications to the soft anomalous dimension
for massive particles at two loops are remarkably simple~\cite{Becher:2009kw,
Ferroglia:2009ep,Chien:2011wz}. It is not inconceivable that they might 
be understood in terms of a deformation of the conformal theory defined 
in the massless case.

In a final speculative remark, we note that the basic gauge-theoretical
ingredients of infrared operators -- form factors, matrix elements of Wilson 
lines and anomalous dimensions -- have definitions that apply in principle 
beyond perturbation theory, and indeed, when strong-coupling data are
available, they have been shown to smoothly match onto the perturbative 
definitions~\cite{Alday:2007hr,Alday:2009zf}. If the dictionary mapping these 
quantities to the celestial sphere can be fully understood, it is possible
that non-perturbative features of the four-dimensional theory could be 
gleaned in the simpler context of the two-dimensional celestial framework.
To this end, a more precise matching between the ideas presented here
and the well-developed framework initiated in~\cite{Strominger:2013lka,
Strominger:2013jfa} will certainly be useful.


\section*{Acknowledgements}
This work was partially supported by the Italian Ministry of University and
Research (MIUR), grant PRIN 20172LNEEZ. I wholeheartedly thank Claude 
Duhr for providing the initial motivation for this project, and for important and 
enlightening discussions. I thank Michele Caselle for sharing his understanding 
of conformal field theories and for several useful insights. I thank Lance Dixon,
Einan Gardi, Marco Meineri, and George Sterman for their reading of the 
manuscript prior to publication, and for useful and significant observations.


\bibliographystyle{JHEP}
\bibliography{ConfIRbib}


\end{document}